\documentclass[referee]{aa}
\usepackage{epsfig}
\begin{document}

\newcommand{\gsim}{\hbox{\rlap{$^>$}$_\sim$}}
  \thesaurus{06;  19.63.1}

\authorrunning{S. Dado, A. Dar \& A. De R\'ujula}
\titlerunning{Afterglow of GRB 010921}
\title{Is there a 1998bw-like supernova in the afterglow of 
gamma ray burst 010921?} 

\author{Shlomo Dado$^{^1}$, Arnon Dar$^{^1}$ and
A. De R\'ujula$^{^2}$}
\institute{1. Physics Department and Space Research Institute, Technion\\ 
               Haifa 32000, Israel\\
           2. Theory Division, CERN, CH-1211 Geneva 23, Switzerland}

\maketitle

\begin{abstract}

We use the very simple and successful
Cannonball Model (CB) of gamma ray bursts (GRBs)
and their afterglows (AGs) to analyze the
observations of the strongly extinct optical AG of GRB
010921 with ground-based telescopes at early times, and with
the HST at later time.
We show that GRB 010921 was indeed associated with a
1998bw-like supernova at the GRB's redshift.

\end{abstract}

The identity of the progenitors of gamma ray bursts is still debated.  It
has been suggested that GRBs are produced by highly relativistic jets
(e.g., Shaviv and Dar 1995, Dar 1998) mostly from supernovae like SN1998bw
(Dar and Plaga 1999, Dar 1999a, Dar and De R\'ujula 2000 and references
therein).  Evidence for a SN1998bw-like contribution to a GRB afterglow
(Dar 1999a; Castro-Tirado \& Gorosabel 1999) was first found by Bloom et
al.~(1999a) for GRB 980326, but the unknown redshift prevented a
quantitative analysis. The AG of GRB 970228 (located at redshift $\rm
z=0.695$) appears to be overtaken by a light curve akin to that of
SN1998bw (located at $\rm z_{bw}=0.0085$), when properly scaled by their
differing redshifts (Dar 1999b; Reichart 1999; Galama et al.~2000).
Evidence of similar associations was found for GRB 990712 (Hjorth et
al.~2000; Sahu et al.~2000;  Bjornsson et al.~2001), GRB 980703 (Holland
et al.~2000), GRB 000418 (Dar and De R\'ujula 2000), GRB 991208
(Castro-Tirado et al.~2001), GRB 970508 (Sokolov et al. 2001) and GRB
000911 (Lazzati et al. 2001). 

Dar and De R\'ujula (2000) and Dado et al.
(2001a) have shown that the optical AG of all relatively nearby GRBs with
known redshift (all the ones with $\rm z<1.12$) contain evidence
or hints for a SN1998bw-like contribution to their optical AG, suggesting
that most ---and perhaps all--- of the long duration GRBs are associated
with 1998bw-like supernovae, but  more cases are needed before a firm 
conclusion can be reached.

On 2001 September 21.21934 UT a bright gamma ray burst (GRB 010921) was 
detected and localized (Ricker et al. 2002) by the High Energy
Transient Explorer (HETE2) and the Interplanetary Network IPN (Hurley et 
al. 2001). Early observations of the error box of GRB 010921 with large 
aperture telescopes found no evidence of an optical afterglow  to the
limit R $\sim$ 20.5 (Fox et al. 2001, Henden et al. 2001). However,  
follow-up
observations by Price et al. (2001) resulted in the discovery of its
optical and radio AGs. This event was the first GRB localized by 
HETE2 which has resulted in the detection of an AG.  Further
observations by various telescopes (Park et al. 2002, Price 2002a and
references therein) have shown that the GRB is located in a relatively 
nearby host galaxy with redshift $\rm z=0.451$ (Djorgovski et al.  
2001) and that 
its optical afterglow exhibits a fast temporal decline and has a large 
spectral slope, $-2.22\pm 0.23$, after correcting for extinction in  
our galaxy,
(E(B-V)=0.148; Schlegel et al. 1998), in the direction of the 
GRB.  Due to its relative proximity and fast temporal decline, the AG of 
GRB 010921 should have shown an SN1998bw-like contribution, after
subtraction of its host galaxy contribution, if it was associated with a 
supernova akin to 1998bw, located at $\rm z=0.451$.

In order to detect or constrain an underlying supernova in GRB 010921,
observations with HST were obtained on 2001 October 26, November 6 and 25,
and 2002 Jan 4 by Price et al. (2002b). From a fireball-model analysis of
these observations, these authors identified a break in the optical
light-curve around day 35 which they argued is due to the collimated
ejecta, and concluded that the existence of a SN with a luminosity greater
than 20\% of SN 1998bw at $\rm z=0.451$ is ruled out. However, in this
note, using the very successful Cannonball Model of GRBs (Dar and De
R\'ujula 2000) and their afterglows (Dado et al. 2001a) to analyze the
light-curve and spectral evolution of the AG of GRB 010921, we show that
there is strong and clear evidence that GRB 010921 was associated with a
supernova akin to 1998bw, at the GRB's redshift\footnote{
Another nearby GRB  that was predicted (e.g., Dado et al. 2001b) to show 
a clear  SN1998bw like bump in its optical AG
is GRB 011121 at z=0.36. Indeed, from their analysis of  
ground based and HST observations on December 4, 2001
Garnavitch et al. (2002) concluded that the AG of GRB 011121
shows the anticipated SN1998bw-like contribution. Price
and coauthors concluded from observations with HST that ``This curious 
bump is inconsistent with an
underlying SN similar to SN 1998bw'' (Bloom et al. 2002) one day, and ``it
appears that the case for an underlying SN in GRB 011121 is well
established'' the day after (Kulkarni et al. 2002).}.

\section*{The CB model of GRBs}
In the CB model, long-duration GRBs and their AGs are produced in core 
collapse supernovae by jets of highly relativistic ``cannonballs'' that 
pierce through the supernova shell. The AG --the persistent radiation in 
the direction of an observed GRB-- has three origins: the ejected CBs, the 
concomitant SN explosion, and the host galaxy. These components are
usually unresolved in the measured ``GRB afterglows'', so that the
corresponding light curves and spectra refer to the cumulative energy flux
density:
\begin{equation}
\rm    F_{AG}=F_{CBs}+F_{SN}+F_{HG} .
\label{sum}
\end{equation}
The contribution of the candidate host galaxy depends on the
angular aperture of the observations and it is usually determined
at late times, when the CB and SN contributions have become
negligible.

Core-collapse supernovae (SN of Types II/Ib/Ic)
are far from being standard candles. But if their explosions
are fairly asymmetric ---as they would be if a fair fraction of
them emit two opposite jets of CBs---  much of the variability could be a  
reflection of the varying angles from which we see their
non-spherically expanding shells.
Exploiting this possibility to its extreme, we use
SN1998bw as an ansatz standard candle (Dado et al. 2001a and references  
therein).
Let the energy flux density of SN1998bw
at redshift $\rm z_{bw}=0.0085$ (Galama et al. 1998)
be $\rm F_{bw}[\nu,t]$.
For a similar SN placed at a redshift $\rm z$:
\begin{eqnarray}
{\rm F_{SN}[\nu,t] = } &&
{\rm{1+z \over 1+z_{bw}}\;
{D_L^2(z_{bw})\over D_L^2(z)}}\, \times\nonumber \\
&&{\rm F_{bw}\left[\nu\,{1+z \over 1+z_{bw}},t\,
{1+z_{bw} \over 1+z}\right]\; A(\nu,z)}\, ,
\label{bw}
\end{eqnarray}
where $\rm D_L(z)$ is the luminosity distance\footnote{The cosmological 
parameters we use in our calculations are:
$\rm H_0=65$ km/(s Mpc), ${\rm \Omega_M}=0.3$ and
${\rm \Omega_\Lambda}=0.7$.}
and $\rm A(\nu,z)$ is the extinction along
the line of sight. The selective extinction in the Galaxy in the direction  
of GRB 010921 is $\rm E(B-V)=0.148$ (Schlegel et al. 1998).
One can use standard conversion of E(B-V) to the
attenuation in our galaxy,
due to dust along the line of sight to the GRB at  
different wave lengths. The  selective extinction in our Galaxy
in the direction of GRB 010921 is expected to increase the
spectral slope of the observed optical light
by $-0.64\pm 0.04$.

The unextinct late time contribution of a CB to the GRB afterglow is  
given by:
\begin{equation}
\rm
F_{CB}=f \; [\gamma(t)]^{3\alpha-1}\;[\delta(t)]^{3+\alpha}\,
\nu^{-\alpha} ,
\label{fluxdensity2}
\end{equation}
where $\rm f$ is a normalization constant (see Dado et al. 2001a
for its theoretical estimate), $\rm \alpha\approx -1.1$ is
the spectral index of the synchrotron radiation from Fermi-accelerated  
electrons in the CB whose acceleration rate is in equilibrium with their  
synchrotron cooling rate,
 $\rm \gamma(t)$ is the Lorentz factor of
the CB, and $\rm \delta(t)$ is its Doppler factor:
\begin{equation}
\rm
\delta\equiv\rm{1\over\gamma\,(1-\beta\cos\theta)}
\simeq\rm {2\,\gamma\over (1+\theta^2\gamma^2)}\; ,
\label{doppler}
\end{equation}
whose approximate expression is valid for small observing angles
$\theta\ll 1$ and $\gamma\gg 1$,
the domain of interest for GRBs.
For an interstellar medium of constant baryon density $\rm n_p$, the  
Lorentz
factor, $\rm\gamma(t)$, as a function of observer's time,
is given by (Dado et al. 2001a):
\begin{eqnarray}
\rm \gamma&=&\rm\gamma(\gamma_0,\theta,x_\infty;t)
=\rm {1\over B} \,\left[\theta^2+C\,\theta^4+{1\over C}\right]\nonumber\\ 
\rm C&\equiv&\rm
\left[{2\over B^2+2\,\theta^6+B\,\sqrt{B^2+4\,\theta^6}}\right]^{1/3}
\nonumber\\
\rm B&\equiv&\rm
{1\over \gamma_0^3}+{3\,\theta^2\over\gamma_0}+
{6\,c\, t\over  (1+z)\, x_\infty}
\label{cubic}
\end{eqnarray}
where $\gamma_0=\gamma(0)$, and
\begin{equation}
\rm
x_\infty\equiv{N_{CB}\over\pi\, R_{max}^2\, n_p}
\label{range}
\end{equation}
characterizes the CB's slow-down in terms of 
$\rm N_{CB}$: its baryon number, and $\rm R_{max}$:
its radius (it takes a distance $\rm x_\infty/\gamma_0$ for
the CB to half its original Lorentz factor).

After a correction for selective extinction in our
galaxy ($\rm E(B-V)=0.148$  magnitudes in the
direction of GRB 010921) the photometric measurements of its
optical AG yield a spectral slope  $\alpha=-2.22\pm 0.23$
(Park et al. 2001; Price et al. 2002a). This slope is much steeper
than the normally observed spectral slopes of GRB afterglows:  $\sim\, 
-0.5$ at early time (the first couple of days)  and $\sim\, -1.1$  
later.  
We interpret this steepening as the effect of selective extinction in  
the host galaxy; a similarlly strong steepening of the spectral index, 
also consistent with strong extinction in the host galaxy,  has
been observed in the AG of a few  GRBs, e.g., GRB 
990712 (Bjornsson et al. 2001) and GRB 000926 (Harrison et al. 2001). 
Supportive 
evidence for strong extinction in the host galaxy of GRB 010921 is
provided by the spectral index of its continuum light
and by the relative intensity of  its measured $\rm H_\alpha$ to
$\rm H_\beta$ emission lines (Price et al. 2002a),
which is different from the expected intrinsic
ratio of 2.87 (Mathis 1983) and yields an optical depth at $\rm H_\beta$ 
due to dust of 1.51. 

\subsection*{Light-curve evidence}
Assuming an intrinsic $\alpha=-1.1$ for the  late-time AG
(in agreement with all observed AGs of GRBs of known
redshift, see Dado et al. 2001a), the best-fitted
parameters to the  AG of GRB 011121 are: $\rm \gamma_0 =
1013$, $\rm \theta = 0.15\, mrad$, and $\rm x_\infty = 0.48\, Mpc$. In
Figs. 1 to 3, we show the fitted light curves and the observations in the
V, R and I bands, assuming a SN1998bw-like contribution at $\rm z=0.451$.
The relative intensities were adjusted to fit the early time data. The
difference between the expected intensity ratios for a spectral slope -1.1
and the observed relative intensities were used to correct the
SN1998bw contribution for selective extinction in the host galaxy and in
the Milky Way. The selective extinction in our Galaxy 
and in the host was also used to estimate roughly the overall 
attenuation normalization: $\rm A_R\approx 2.8$ magnitudes in the R band. 
The fits are clearly very satisfactory. 

\subsection*{Spectral evidence}

The spectral index of the optical AG for most GRBs after a couple of 
days is approximately $-1.1$ (e.g., Simon et al., 2001).
When dominated by a supernova contribution, the optical
afterglows of GRBs become much more red 
(see, e.g., Reichart 1999; Bloom et al. 1999). 
The spectral index 
of the optical AG of GRB 010921  steepened 
to $-3.5\pm 0.3$ on day 35 and to $ -4.5$ on 
day 46 (Price et al. 2002b ) as expected
when the AG is dominated by a supernova
like SN1998bw. This is demonstrated in Figs. 4 and 5 where we compare the 
colours of SN1998bw at $\rm z=0.451$ ---after extinction in the host galaxy
and in ours--- and the
colours of the AG of GRB 010921, as measured with HST on days 35 and 46 
after burst. The colours observed by HST  are in good agreement with the
expectation.

\section*{Conclusions}

Contrary to the questionable\footnote{See,
for instance, the chapter ``Uses and abuses of  fireballs'' in Dado et  
al. 2001a, and the honest and devastatingly
section ``Open issues and problems'' in Ghisellini  
2001.} and inconclusive fireball-model analysis
of the optical afterglow of GRB 010921, our analysis,
 based on the Cannonball Model, shows  strong evidence for
the contribution of a supernova akin to SN1998bw, seen at
the GRB's redshift.  The evidence is clearly there in all of the
relevant observations:
the I, R and V light curves, and the spectra at the times when the
supernova's contribution is dominant.

\newpage
\begin{figure}[]
\hskip 2truecm
\vspace*{2cm}
\hspace*{-1.7cm}
\epsfig{file=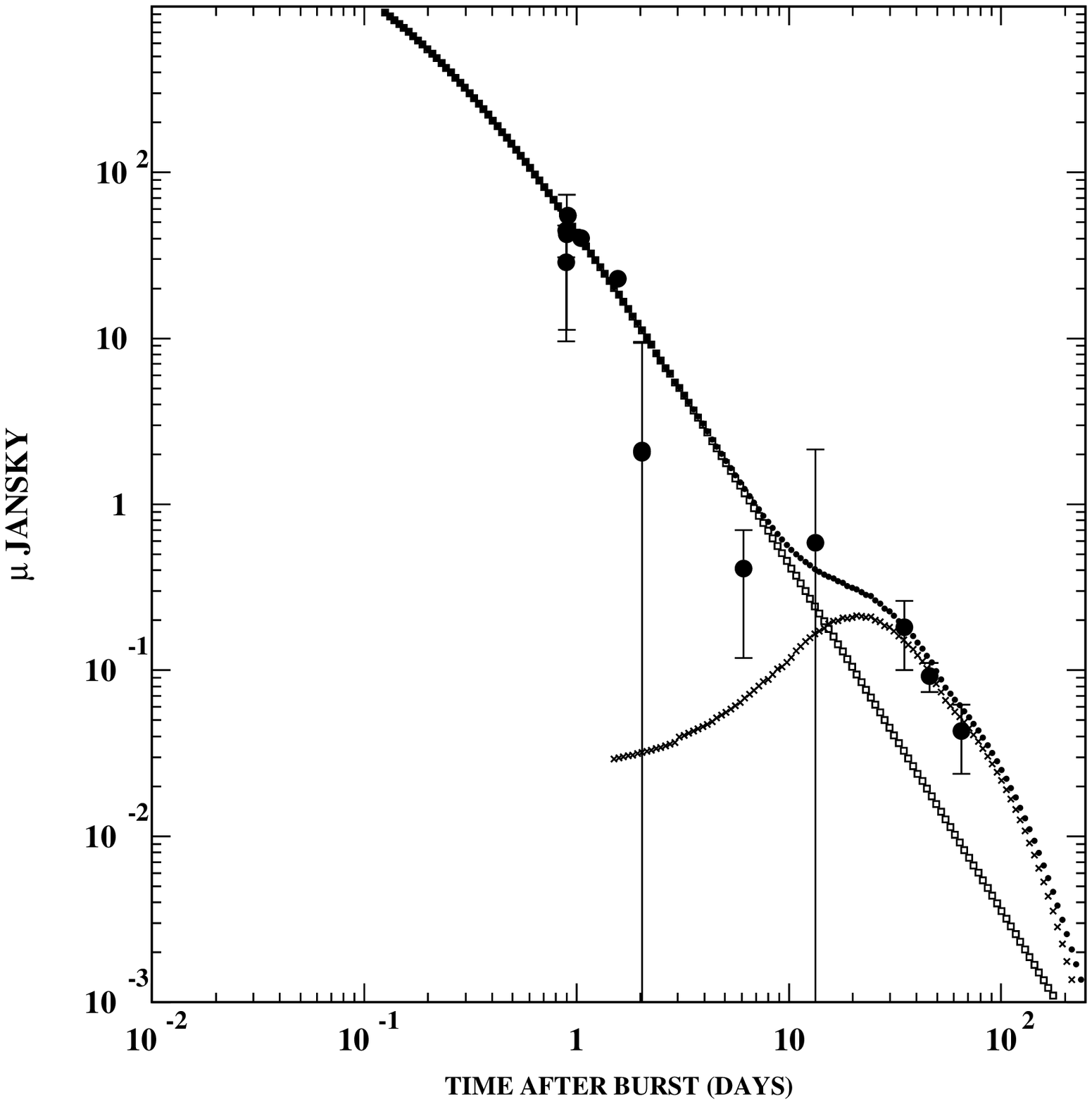, width=16cm}
\caption{ The CB model fit and and the observations (Park et al 2001;
Price et al. 2002a, 2002b) of the R band afterglow of GRB 010921 at $\rm 
z=0.45$. The CB's AG (the line of squares) is given by Eqs.~(3) to (5).
The contribution from a 1998bw-like supernova placed at the GRB's
redshift and modified by galactic extinction, Eq.~(\ref{bw}),
is indicated by a line of crosses.
The SN 1998bw-like contribution, is clearly observed.}
\label{fig1121}
\end{figure}
\newpage

\begin{figure}[]
\hskip 2truecm
\vspace*{2cm}
\hspace*{-1.7cm}
\epsfig{file=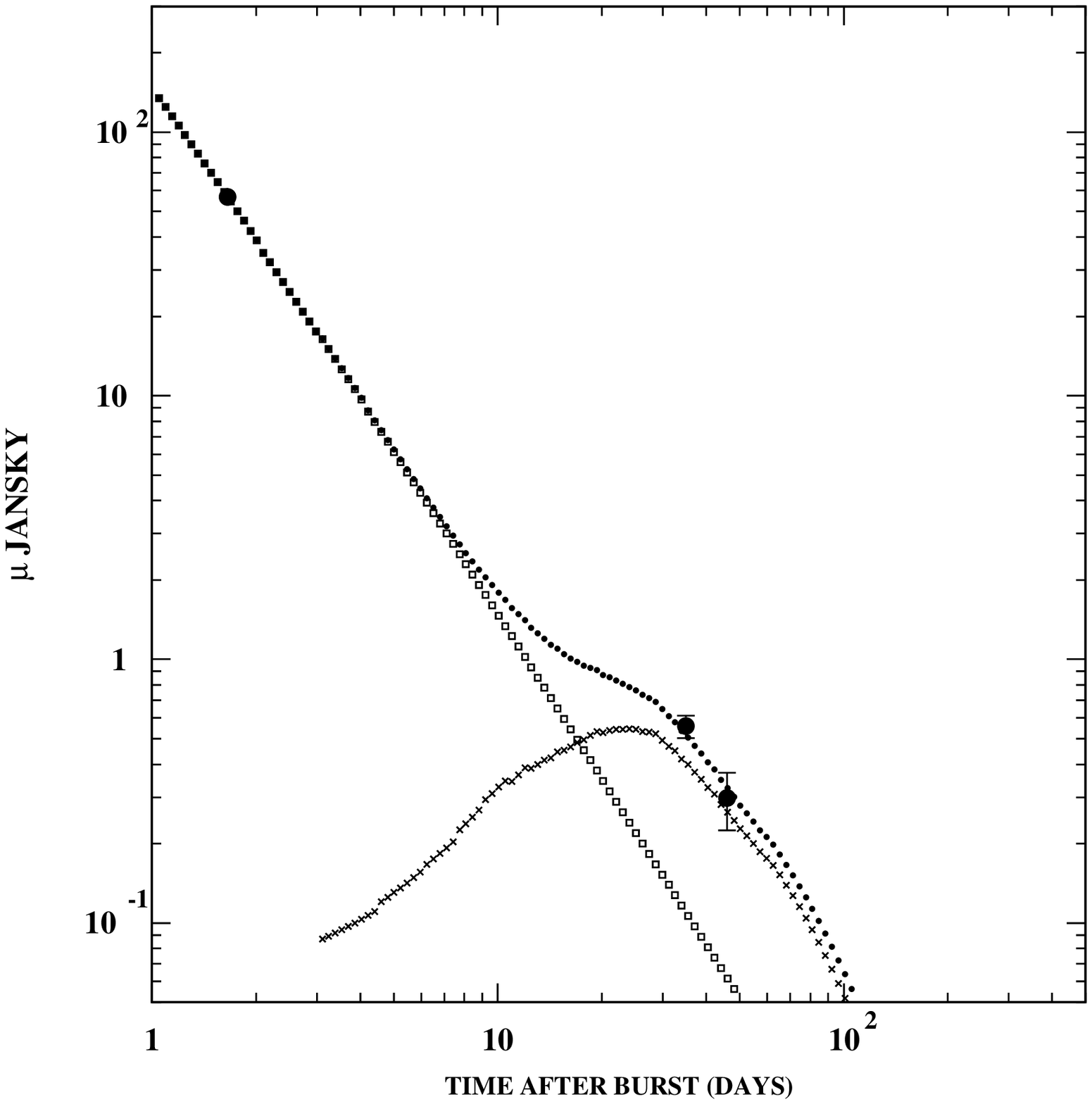, width=16cm}
\caption{ The CB model fit and and the observations (Park et al 2001;
Price et al. 2002a, 2002b) of 
the I band afterglow of GRB 010921 at $\rm z=0.451$.
The CB's AG (the line of squares) is given by Eqs.~(3) to (5).
The contribution from a 1998bw-like supernova placed at the GRB's
redshift and modified by galactic extinction, Eq.~(\ref{bw}),
is indicated by a line of crosses.
The SN 1998bw-like contribution, is clearly observed.}
\label{fig1121}
\end{figure}
\newpage

\begin{figure}[]
\hskip 2truecm
\vspace*{2cm}
\hspace*{-1.7cm}
\epsfig{file=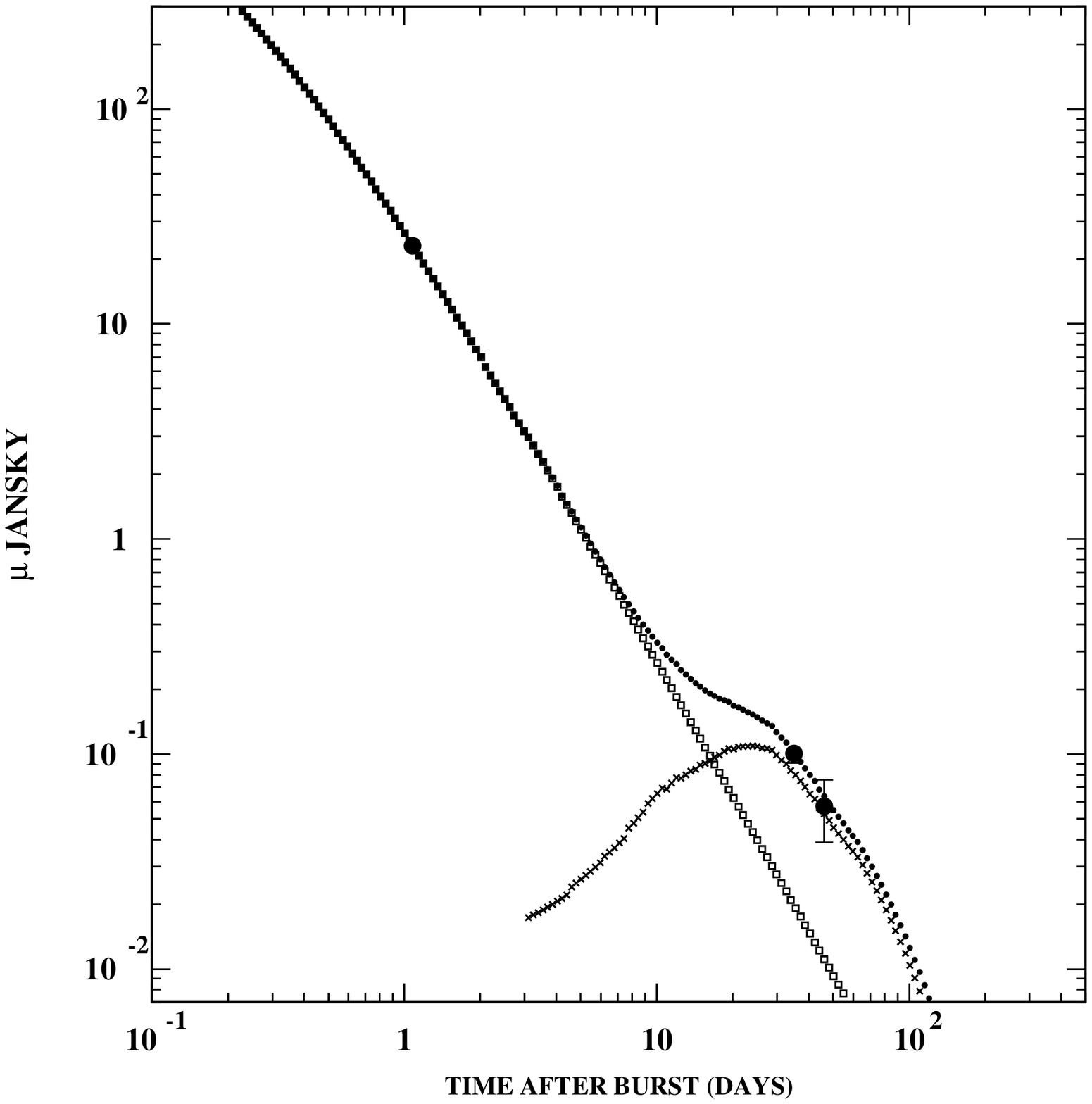, width=16cm}
\caption{ The CB model fit and and the observations
(Park et al. 2001; Price et al. 2002a, 2002b)
of the V band afterglow of GRB 010921 at $\rm z=0.451$.
The CB's AG (the line of squares) is given by Eqs.~(3) to (5).
The contribution from a 1998bw-like supernova placed at the GRB's
redshift and modified by galactic extinction, Eq.~(\ref{bw}),
is indicated by a line of crosses.
The SN 1998bw-like contribution, is clearly observed.}
\label{fig1121}
\end{figure}
\newpage

\begin{figure}[]
\hskip 2truecm
\vspace*{2cm}
\hspace*{-1.7cm}
\epsfig{file=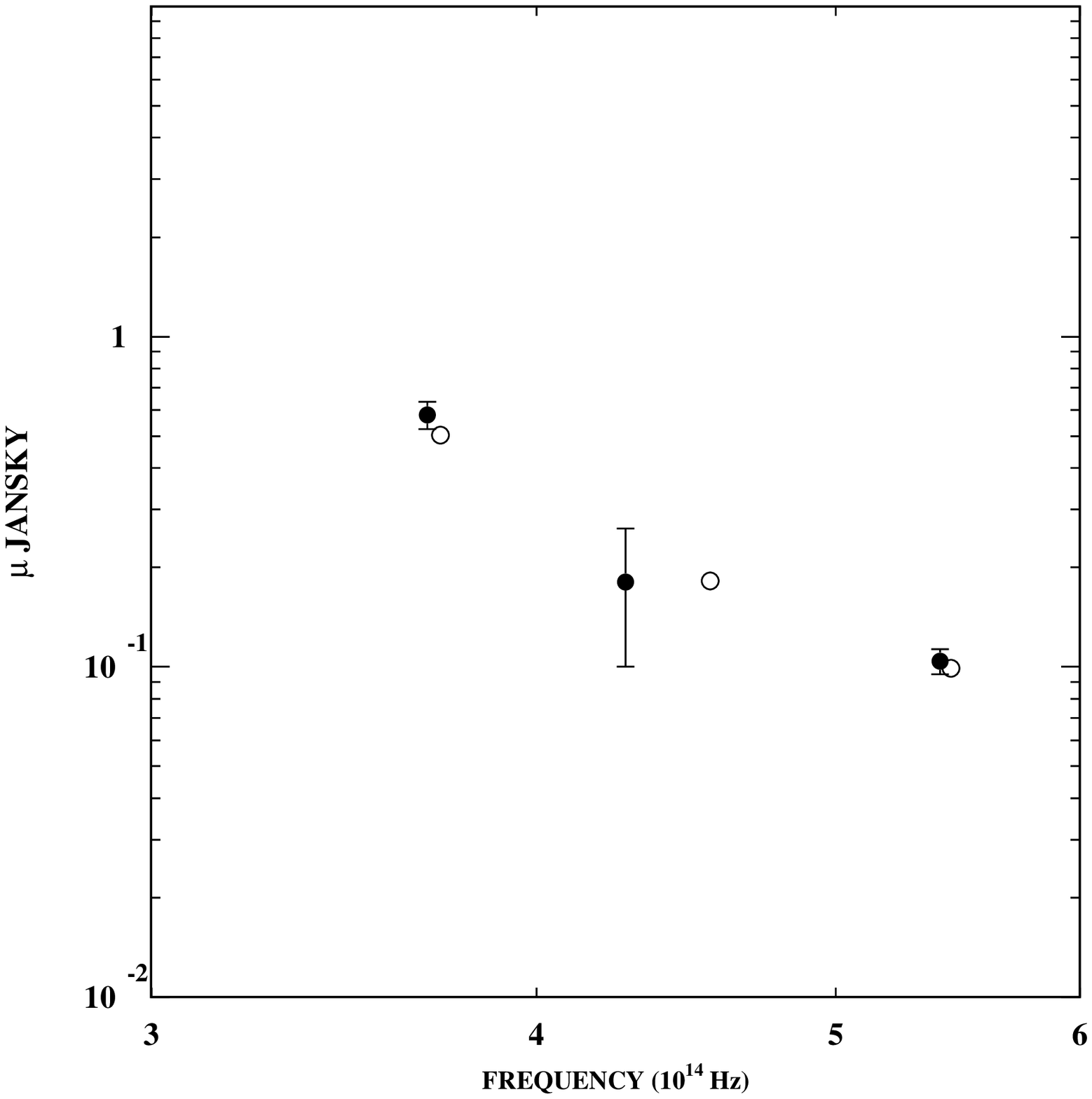, width=16cm}
\caption{ Comparison between the CB model predictions (circles) for the 
spectral energy density in the I, R, V 
bands  of the afterglow of GRB 010921 on day 35 after burst 
and the measurements (full circles) with HST (Price et al. 2002b). 
The  predicted values  are the sum of the light from the cannonballs and 
from a SN1998bw placed at $\rm z=0.451$, corrected for extinction in 
the host galaxy and in ours.}
\label{fig1121c35}
\end{figure}
\newpage
\begin{figure}[]
\hskip 2truecm
\vspace*{2cm}
\hspace*{-1.7cm}
\epsfig{file=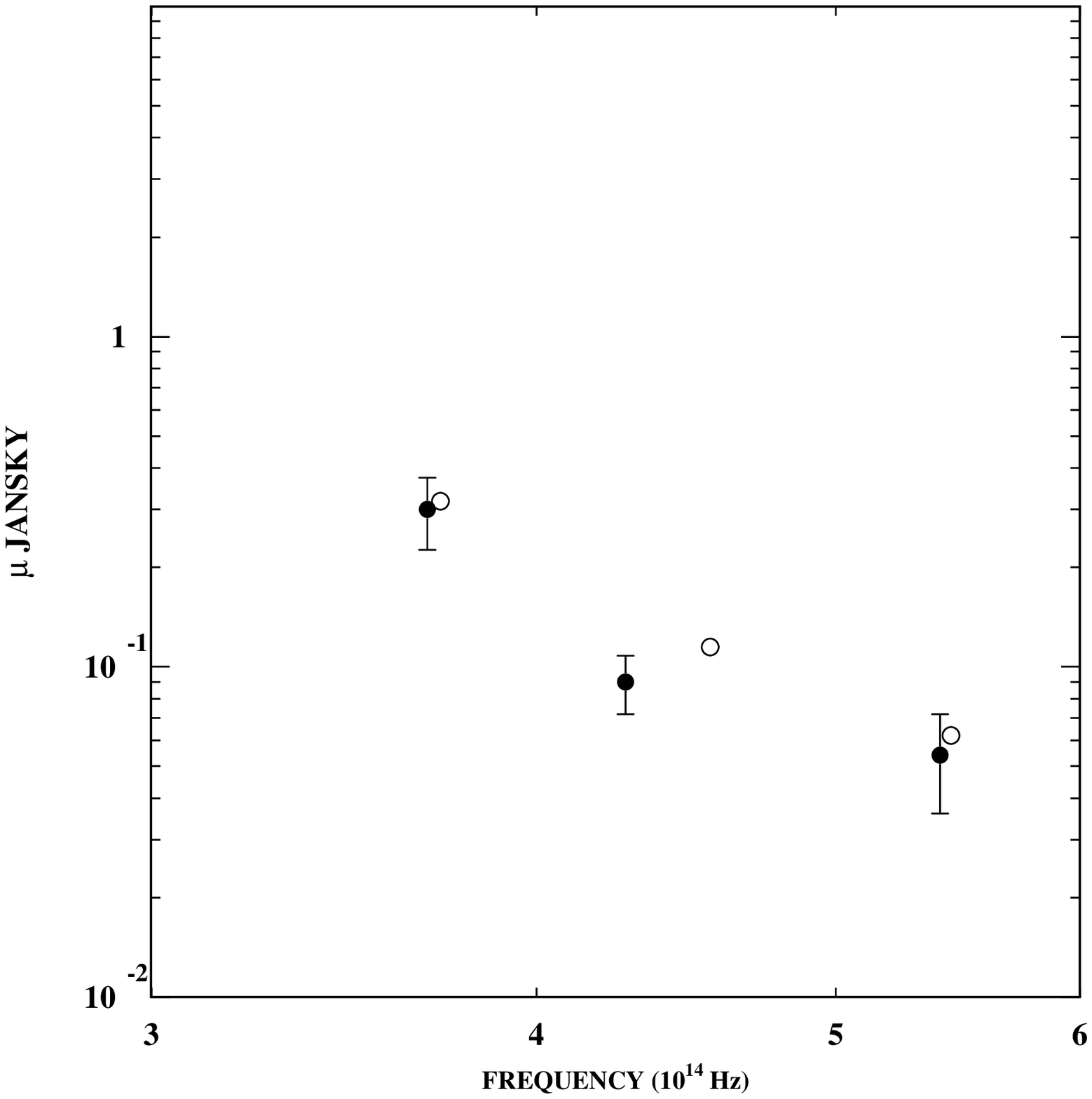, width=16cm}
\caption{ Comparison between the CB model predictions (circles) for the 
spectral energy density in the I, R, V 
bands  of the afterglow of GRB 010921 on day 46 after burst 
and the measurements  
(full circles) with HST (Price et al. 2002b).
The  predicted values  are the sum of the light from the CB  and 
from a SN1998bw placed at $\rm z=0.451$, corrected for extinction in 
the host galaxy and in ours.}
\label{fig1121c46}
\end{figure}
\newpage
\end{document}